\documentclass[12pt]{article}
\usepackage{epsfig}

\usepackage{a4}
\usepackage{a4wide}

\def\gs{\mathrel{
   \rlap{\raise 0.511ex \hbox{$>$}}{\lower 0.511ex \hbox{$\sim$}}}}
\def\ls{\mathrel{
   \rlap{\raise 0.511ex \hbox{$<$}}{\lower 0.511ex \hbox{$\sim$}}}}

\newcommand{\ba}{\begin{array}{c}}
\newcommand{\baz}{\begin{array}{cc}}
\newcommand{\bad}{\begin{array}{ccc}}
\newcommand{\bav}{\begin{array}{cccc}}
\newcommand{\bea}{\begin{equation} \begin{array}{c}}
\newcommand{\eea}{ \end{array} \end{equation}}
\newcommand{\ea}{\end{array}}

\newcommand{\be}{\begin{eqnarray}}
\newcommand{\ee}{\end{eqnarray}}

\begin{document}

\begin{titlepage}
\title{\vspace*{-2.0cm}
\bf\Large
$\eta_{Q}$ meson photoproduction in ultrarelativistic heavy ion collisions
\\[5mm]\ }

\author{
Gong-Ming Yu$^{1}$\thanks{Email: \tt ygmanan@163.com},~~~Gao-Gao Zhao$^{2}$,~~~Zhen Bai$^{1}$,~~~Yan-Bing Cai$^{2}$\thanks{Email: \tt myparticle@163.com},\\ Hai-Tao Yang$^{3}$\thanks{Email: \tt yanghaitao205@163.com},~~~and~Jian-Song Wang$^{1}$\thanks{Email: \tt jswang@impcas.ac.cn}
\\ \\
{\normalsize \it $^{1}$CAS Key Laboratory of High Precision Nuclear Spectroscopy and Center for Nuclear} \\
{\normalsize \it Matter Science, Institute of Modern Physics, Chinese Academy of Sciences,}\\
{\normalsize \it Lanzhou 730000, People's Republic of China}\\
{\normalsize \it $^2$ Department of physics, Yunnan University, Kunming 650091, People's Republic of China}\\
{\normalsize \it $^3$ School of physics and electronic information engineering, Zhaotong University,}\\
{\normalsize \it Zhaotong 657000, People's Republic of China}
}
\date{}
\maketitle
\thispagestyle{empty}

\begin{abstract}
\noindent
The transverse momentum distributions for inclusive $\eta_{c,b}$ meson described by gluon-gluon interactions from photoproduction processes in relativistic heavy ion collisions are calculated. We considered the color singlet (CS) and color octet (CO) components with the framework of non-relativistic Quantum Chromodynamics (NRQCD) into the production of heavy quarkonium. The phenomenological values of the matrix elements for the color-singlet and color-octet components give the main contribution to the production of heavy quarkonium from the gluon-gluon interaction caused by the emission of additional gluon in the initial state. The numerical results indicate that the contribution of photoproduction processes cannot be negligible for mid-rapidity in p-p and Pb-Pb collisions at the Large Hadron Collider (LHC) energies.

\end{abstract}

\end{titlepage}

\section{\label{sec:intro}Introduction}

Heavy quarkonium is a multiscale system which can probe all regimes of Quantum Chromodynamics (QCD), and present an ideal laboratory for testing the interplay between perturbative and nonperturbative QCD within a controlled environment. In the recent years, many measurement reports have been published by ALICE collaboration\cite{1,2}, CMS collaboration\cite{3,4}, ATLAS collaboration\cite{5,6}, and LHCb collaboration\cite{7,8} at the Large Hadron Collider (LHC) energies; several theoretical approaches have been proposed such as the color-singlet (CS) mechanism\cite{9,10}, the color-octet (CO) mechanism\cite{11,12}, the color evaporation mechanism\cite{13,14}, the color-dipole mechanism\cite{15,16,17,18}, the mixed heavy-quark hybrids mechanism\cite{19}, the recombination mechanism\cite{19x0,19x1,20,21,22}, the photoproduction mechanism\cite{23,24,25,26,27,28,29}, the potential non-relativistic Quantum Chromodynamics (pNRQCD) approach\cite{30,31,32}, the transverse-momentum-dependent factorization approach\cite{33}, the transport approach\cite{33x0,33x1,33x2,34,35,36}, the $k_{T}$-factorization approach\cite{37,38,39,40}, the fragmentation approach\cite{41,42,43,44,45,46,47}, and the non-relativistic Quantum Chromodynamics (NRQCD) approach\cite{48,49,50,51,52,53,54,55,56,57,58,59,60,61,62}. Among them, the NRQCD approach, which takes into account contributions of color-singlet component and color-octet components with the nonperturbative long-distance matrix elements (LDME), is the most successful in phenomenological studies. The long-distance matrix elements are process-independent, and can be classified in terms of the relative velocity for the heavy quarks in the bound state. But, the heavy quarkonium production mechanism is still not fully understood.

In this study, we extend the hard photoproduction mechanism\cite{63} to the heavy quarkonium production and investigate the production of inclusive $\eta_{c,b}$ meson in p-p and Pb-Pb collisions at the LHC. According to Ref. \cite{64}, the light $q\bar{q}$ contributions for heavy quarkonium production are negligible, therefore in this work we only consider the contributions of gluon-gluon processes caused by the emission of additional gluons, that is different from our previous work\cite{24} based on the method developed in Ref. \cite{63x0,63x1}. In high-energy collisions, the partons from the nucleus can emit high-energy photons that can fluctuate into gluons and then interact with the partons of the other nucleus. Hence we consider that the hard photoproduction processes of a charged parton of the incident nucleon can emit a high energy photon in high energy nucleus-nucleus collisions.

The paper is organized as follows. In section 2 we present the photoproduction of inclusive $\eta_{c,b}$ from gluon-gluon interactions at LHC. The numerical results for large-$p_{T}$ $\eta_{c,b}$ meson production in p-p collisions and Pb-Pb collisions at LHC are given in section 3. Finally, the conclusion is given in section 4.

\section{\label{sec:LowLim}General formalism}

In relativistic heavy ion collisions, the production of $\eta_{Q}$ mesons by the gluon-gluon (g-g) processes from the initial parton interaction can be divided into three processes: direct g-g processes, semi-elastic resolved photoproduction and inelastic resolved photoproduction processes.

In direct processes, the parton (gluon) $a$ of the incident nucleus $A$ interacts with the parton (gluon) $b$ of another incident nucleus $B$ by the interaction of $gg\rightarrow \eta_{c}g$. The invariant cross section of large-$p_{T}$ $\eta_{Q}$ meson of the process $(A + B \rightarrow \eta_{Q} + X)$ is described in the pQCD parton model on the basis of the factorization theorem and can be written as
\begin{eqnarray}
\frac{d\sigma^{LO}_{A  B \rightarrow \eta_{Q}  X}}{dp_{T}^{2}dy}=\int dx_{a}f_{g/A}(x_{a},Q^{2})f_{g/B}(x_{b},Q^{2})\frac{x_{a}x_{b}}{x_{a} - x_{1}}\frac{d\hat{\sigma}}{d\hat{t}}(gg\rightarrow \eta_{Q}g),
\end{eqnarray}
where the variables $x_{a}$ and $x_{b}=(x_{a}x_{2}-\tau)/(x_{a}-x_{1})$ are the momentum fractions of the partons, $z_{c}$ is the momentum fraction of the final charmed-meson, $x_{1}=\frac{1}{2}(x_{T}^{2}+4\tau)^{1/2}\exp(y)$, $x_{2}=\frac{1}{2}(x_{T}^{2}+4\tau)^{1/2}\exp(-y)$, $x_{T}=2p_{T}/\sqrt{s}$, $\tau=(M/\sqrt{s})^{2}$, and $M$ is the mass of the $\eta_{Q}$ meson;  $f_{g/A}(x_{a},Q^{2})$ and $f_{g/B}(x_{b},Q^{2})$ are the parton distribution functions (PDF) for the colliding partons $a$ and $b$ carrying fractional momentum $x_{a}$ and $x_{b}$ in the interacting nucleons\cite{65},
\begin{eqnarray}
f_{g/A}(x,Q^{2})=R_{A}(x,Q^{2})f_{g}(x,Q^{2}),
\end{eqnarray}
where $R_{A}(x,Q^{2})$ is the nuclear modification factor\cite{66}, and $f_{g}(x,Q^{2})$ is the gluon distribution function of nucleon.

According to NRQCD scaling rules\cite{67,68}, the color-singlet as well as $S$-wave and $P$-wave color-octet components give the main contributions to the production process under consideration\cite{69}
\begin{eqnarray}
\frac{d\hat{\sigma}}{d\hat{t}}(gg\rightarrow \eta_{Q}g)\!\!\!\!&=&\!\!\!\!  |R(0)|^{2} \frac{d\hat{\sigma}}{d\hat{t}}(gg\rightarrow Q\bar{Q}[^{1}S_{0}^{[1]}g])\nonumber\\[1mm]
&&\!\!\!\! +  \langle O_{S}\rangle\frac{d\hat{\sigma}}{d\hat{t}}(gg\rightarrow Q\bar{Q}[^{3}S_{1}^{[8]}g])\nonumber\\[1mm]
&& \!\!\!\!+  \langle O_{P}\rangle\frac{d\hat{\sigma}}{d\hat{t}}(gg\rightarrow Q\bar{Q}[^{1}P_{1}^{[8]}g]).
\end{eqnarray}
The subprocesses cross section of $[^{1}S_{0}^{[1]}]$, $[^{3}S_{1}^{[8]}]$, and $[^{1}P_{1}^{[8]}]$ state are respectively given by\cite{70,70x0}
\begin{eqnarray}
\frac{d\hat{\sigma}}{d\hat{t}}(gg\rightarrow Q\bar{Q}[^{1}S_{0}^{[1]}]g) = \frac{\pi \alpha_{s}^{3}}{\hat{s}^{2}M}\frac{P^{2}}{Q(Q - M^{2}P)^{2}} [(P - M^{4})^{2} + 2M^{2}Q],
\end{eqnarray}
\begin{eqnarray}
\frac{d\hat{\sigma}}{d\hat{t}}(gg\rightarrow Q\bar{Q}[^{3}S_{1}^{[8]}]g)= \frac{\pi \alpha_{s}^{3}}{3M\hat{s}^{2}}\frac{(P^{2} - M^{2}Q)(19M^{4} - 27P)}{M^{2}(Q - M^{2}P)^{2}},
\end{eqnarray}
\begin{eqnarray}
&&\!\!\!\!\!\!\!\!\!\!\frac{d\hat{\sigma}}{d\hat{t}}(gg\rightarrow Q\bar{Q}[^{1}P_{1}^{[8]}]g)=\frac{2\pi \alpha_{s}^{3}}{M^{3}\hat{s}^{2}}\frac{1}{Q(Q - M^{2}P)^{3}}[179M^{4}Q^{3}+ 217M^{10}Q^{2}- 27M^{2}P^{5} + 54M^{6}P^{4}\nonumber\\[1mm]
&&\!\!\!\!\!\!\!\!\!\!   - 27M^{10}P^{3} \!+\! 135PQ^{3} \!+\! 103M^{2}P^{2}Q^{2} \!-\! 212M^{6}PQ^{2}\!-\! 124M^{8}P^{2}Q \!+ \!43M^{12}PQ \!+ \!27P^{4}Q],
\end{eqnarray}
where $M^{2}=\hat{s}+\hat{t}+\hat{u}$, $P=\hat{s}\hat{t}+\hat{t}\hat{u}+\hat{u}\hat{s}$, and $Q=\hat{s}\hat{t}\hat{u}$. Here $\hat{s}$, $\hat{t}$, and $\hat{u}$ are the Mandelstam variables. $R(0)=[M_{H}^{2}\Gamma(H\rightarrow e^{+}e^{-})/4\alpha^{2}e_{Q}^{2}]^{1/2}$ is the wave function value of $\eta_{Q}$ meson for the color-singlet state at the origin\cite{70x1,70x2,70x3,70x4,70x5,70x6}, where $M_{H}\approx2m_{Q}$ is the mass of the heavy quark pairs. The LDMEs of the color-octet components are used as follows
\begin{eqnarray}
&& \langle O_{S}\rangle = \langle R_{\eta_{Q}}[^{3}S_{1}^{(8)}]\rangle = \frac{\pi}{6}\langle0|O_{8}^{\eta_{Q}}[^{3}S_{1}]|0\rangle,\nonumber\\[1mm]
&& \langle O_{P}\rangle = \langle R_{\eta_{Q}}[^{1}P_{1}^{(8)}]\rangle = \frac{\pi}{18}\langle0|O_{8}^{\eta_{Q}}[^{1}P_{1}]|0\rangle.
\end{eqnarray}
For the $\eta_{c}$ meson they are\cite{51}
\begin{eqnarray}
&&|R_{cc}(0)|^{2}\approx0.58\,\mathrm{GeV}^{3},\nonumber\\[1mm]
&& 1.5\times 10^{-3}\, \mathrm{GeV}^{3} < \langle O^{\eta_{c}}_{S}\rangle < 5.3\times10^{-3}\, \mathrm{GeV}^{3},\nonumber\\[1mm]
&& \langle O^{\eta_{c}}_{P}\rangle =\frac{\pi}{18}\times3\times\langle O^{J/\psi}(^{3}P_{0}^{[8]})\rangle\nonumber\\[1mm]
&&\quad\quad\,\,\,= \frac{\pi}{6}\times m_{c}^{2}\times(1.7\pm0.5)\times10^{-2} \,\mathrm{GeV}^{3},
\end{eqnarray}
and for $\eta_{b}$ meson they are\cite{71,72}
\begin{eqnarray}
&& |R_{bb}(0)|^{2}\approx5.3\,\mathrm{GeV}^{3},\nonumber\\[1mm]
&& \langle O_{S}^{\eta_{b}}\rangle\approx0.01\,\mathrm{GeV}^{3},\nonumber\\[1mm]
&&\langle O_{P}^{\eta_{b}}\rangle = \frac{\pi}{18}\times3\times\langle O_{8}^{\gamma(1s)}(^{3}P_{0})\rangle\nonumber\\[1mm]
&&\quad\quad\,\,\,\,\,= \!\frac{5\pi}{6}\!\times m_{b}^{2}\!\times\!(0.0121\pm0.040)\,\mathrm{GeV}^{3},
\end{eqnarray}
where $m_{c}$ ($m_{b}$) is the mass of charm (bottom).

In the semi-elastic resolved photoproduction g-g processes, the parton (gluon) $a$ from resolved photon of the incident nucleus $A$ interacts with the parton (gluon) $b$ of another incident nucleus $B$, and the cross section is given by
\begin{eqnarray}
\frac{d\sigma^{semi.}_{A  B \rightarrow \eta_{Q}  X}}{dp_{T}^{2}dy}= \int dx_{a} dx_{b} f_{\gamma/N}(x_{a}) f_{g/\gamma}(z_{a},Q^{2})f_{g/B}(x_{b},Q^{2}) \frac{x_{a}x_{b}z_{a}}{x_{a}x_{b} - x_{a}x_{2}}\frac{d\hat{\sigma}}{d\hat{t}}(gg\rightarrow \eta_{Q}g),
\end{eqnarray}
where $f_{\gamma/N}(x_{a})$ is the photon spectrum of the nucleus, and $f_{g/\gamma}(z_{a},Q^{2})$ is the parton distribution function of the resolved photon\cite{73}.

\begin{figure}
\begin{center}
\epsfig{file=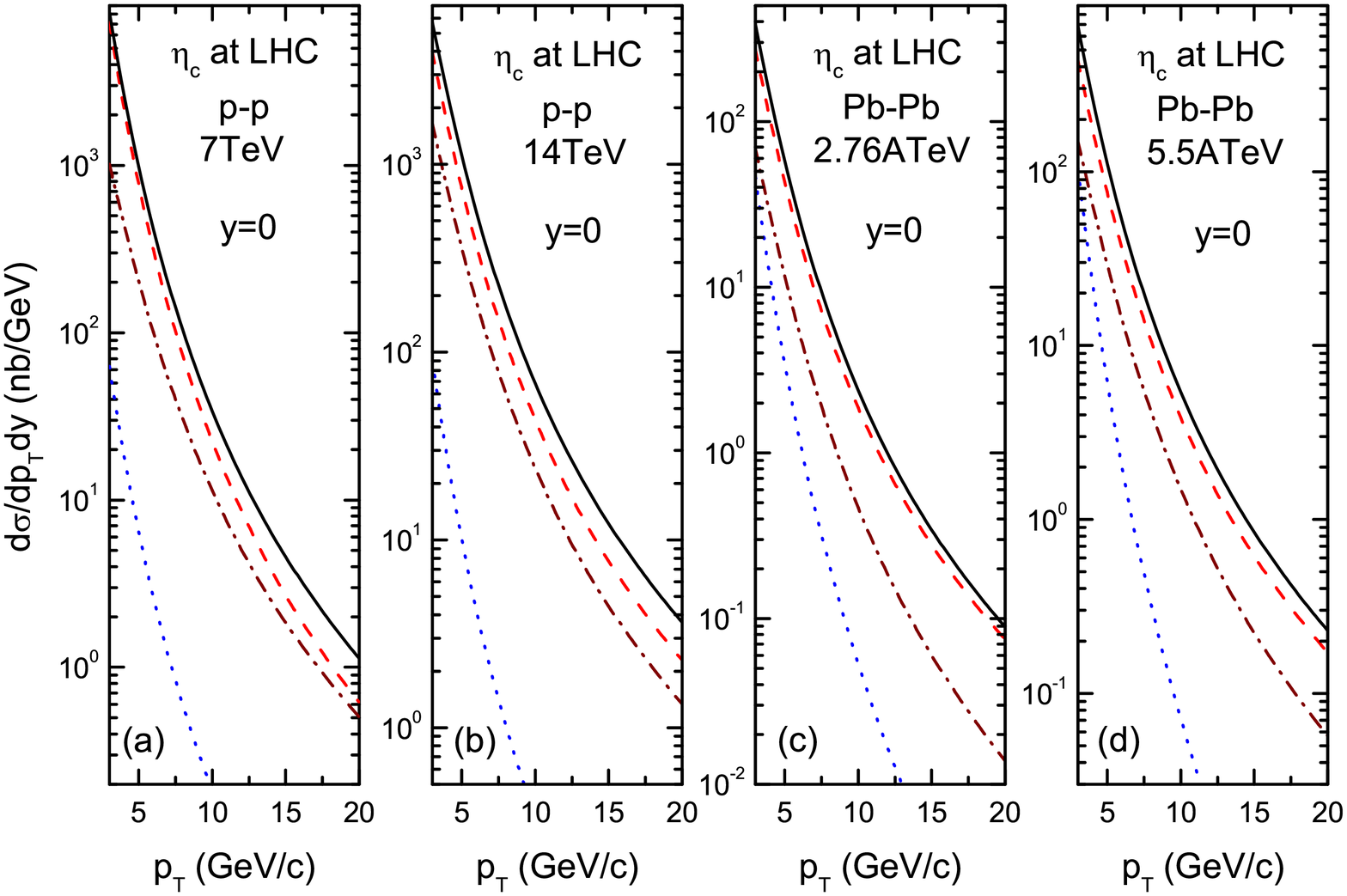,width=14cm,height=6cm}
\caption{\label{fg1}The invariant cross section of large-$p_{T}$ $\eta_{c}$ meson production from gluon-gluon interaction at mid-rapidity in p-p collisions ($\sqrt{s}=7.0 \,\mathrm{TeV}$ and $\sqrt{s}=14.0 \,\mathrm{TeV}$) and Pb-Pb collisions ($\sqrt{s}=2.76 \,\mathrm{TeV}$ and $\sqrt{s}=5.5\, \mathrm{TeV}$) at the LHC. The dashed line (red line) is for the initial gluon-gluon interaction (LO), the dotted line (blue line) for the semielastic hard photoproduction g-g processes (semi.), the dashed-dotted line (wine line) for the inelastic hard photoproduction g-g processes (inel.), and the solid line (black line) for the sum of the above processes.}
\end{center}
\end{figure}

For p-p collisions, the photon spectrum function of a proton can be written as\cite{74,75,76}
\begin{eqnarray}
f_{\gamma/p}(x)=  \frac{\alpha}{2\pi x}[1 + (1 - x)^{2}] \bigg[\ln A_{p} - \frac{11}{6} + \frac{3}{A_{p}} - \frac{3}{2A_{p}^{2}} + \frac{1}{3A_{p}^{3}}\bigg],
\end{eqnarray}
where $x$ is the momentum fraction of photon, $A_{p}=1 + 0.71\,\mathrm{GeV}^{2}/Q_{min}^{2}$ with
\begin{eqnarray}
Q_{min}^{2}= - 2m_{p}^{2} + \frac{1}{2s}\bigg[(s + m_{p}^{2})(s -xs + m_{p}^{2})- (s - m_{p}^{2})\sqrt{(s - xs - m_{p}^{2})^{2}-4m_{p}^{2}xs}\bigg].
\end{eqnarray}
Here $m_{p}$ is the mass of the proton and at high energies $Q_{min}^{2}$ is given to a very good approximation by $m_{p}^{2}x^{2}/(1 - x)$.

For Pb-Pb collisions, the photon spectrum obtained from a semiclassical description of high energy electromagnetic collisions for low photon energies is given by\cite{77,78}
\begin{eqnarray}
f_{\gamma/N}= \frac{2Z^{2}\alpha}{\pi\omega}\ln\bigg(\frac{\gamma}{\omega R}\bigg),
\end{eqnarray}
where $\omega$ is the photon energy, and $R=b_{min}$ is the nucleus radius.

In inelastic resolved photoproduction g-g processes, the parton (gluon) $a^{'}$ from resolved photon which emitted by the charged parton $a$ of the incident nucleus $A$ interacts with the parton (gluon) $b$ of another incident nucleus $B$, and the expression of the cross section is given by
\begin{eqnarray}
\frac{d\sigma^{inel.}_{A  B \rightarrow \eta_{Q}  X}}{dp_{T}^{2}dy}\!\!\!\! &=&\!\!\!\!\int dx_{a} dx_{b} dz_{a} f_{q/A}(x_{a},Q^{2})f_{\gamma/q}(z_{a})f_{g/\gamma}(z'_{a},Q_{\gamma}^{2}) \nonumber\\[1mm]
&&\!\!\!\! \times  f_{g/B}(x_{b},Q^{2})\frac{x_{a}x_{b}z_{a}z'_{a}}{x_{a}x_{b}z_{a} - x_{a}z_{a}x_{2}}\frac{d\hat{\sigma}}{d\hat{t}}(gg\rightarrow \eta_{Q}g),
\end{eqnarray}
where $f_{\gamma/q}(z)$ is the photon spectrum from the charged parton of the incident nucleus. In relativistic hadron-hadron and nucleus-nucleus collisions\cite{64} we have,
\begin{eqnarray}
f_{\gamma /q}(x) = \frac{\alpha }{\pi }e_{Q}^{2}\bigg\{ \frac{1\! +\! (1\! -\! x)^{2}}{x}\bigg(\ln\frac{E}{m} \!-\! \frac{1}{2}\bigg)\!+\! \frac{x}{2}\bigg[\ln\bigg(\frac{2}{x} \!-\! 2\bigg) \!+\! 1\bigg]\!+\! \frac{(2 \!-\! x)^{2}}{2x}\ln \bigg(\frac{2 \!- \!2x}{2 \!-\! x}\bigg)\bigg\},
\end{eqnarray}
with $x$ being the photon momentum fraction.

\begin{figure}
\begin{center}
\epsfig{file=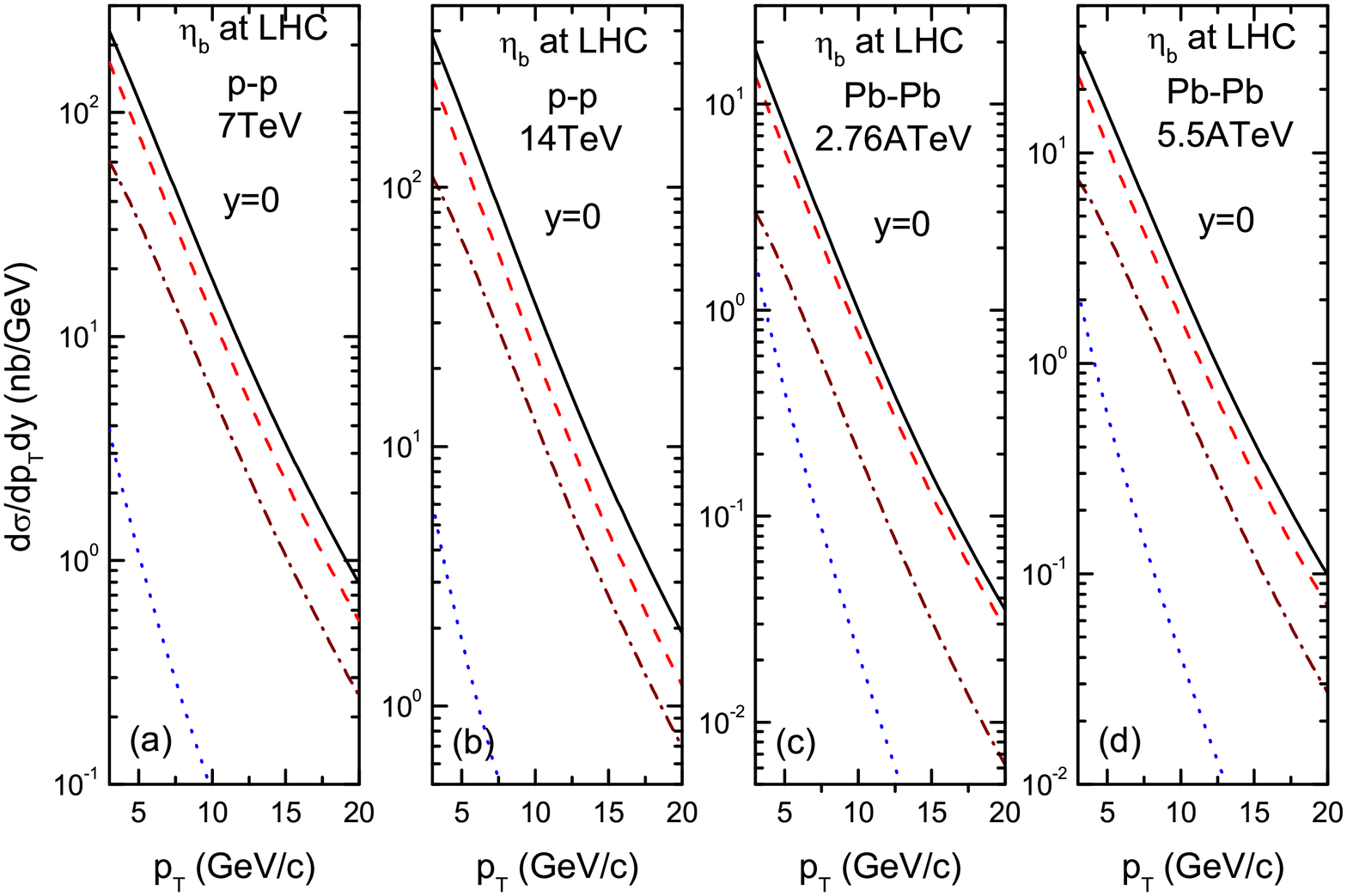,width=14cm,height=6cm}
\caption{\label{fg2}The same as Fig.\,1 but for large-$p_{T}$ $\eta_{b}$ meson production from gluon-gluon interaction at mid-rapidity in p-p and Pb-Pb collisions at the LHC. }
\end{center}
\end{figure}

\section{Numerical results}

In ultrarelativistic high energy nucleus-nucleus collisions, the equivalent photon spectrum obtained with a semiclassical description of high-energy electromagnetic collisions for the nucleus is $f_{\gamma/N}\propto
Z^{2}\ln\gamma$. At LHC energies, the Lorentz factor $\gamma=E/m_{N}=\sqrt{s_{NN}}/2m_{N}\gg1$ becomes very important. Indeed, the equivalent photon spectrum function with Weizs$\ddot{a}$cker-Williams approximation for the proton is $f_{\gamma/p}\propto\ln A\propto\ln(s_{NN}/m_{p}^{2})$, where $m_{p}$ is the proton mass and $\sqrt{s_{NN}}$ is the centre-of-mass energy per nucleon pair. Since $\sqrt{s_{NN}}$ is very high, the photon spectrum function becomes very large. Therefore the contribution of $\eta_{Q}$ meson produced by semielastic hard photoproduction g-g processes cannot be negligible at LHC energies. For the inelastic photoproduction processes, the equivalent photon spectrum function of the charged parton is $f_{\gamma/q}\propto\ln(E/m_{q})=\ln(\sqrt{s_{NN}}/2m_{q})+\ln(x)$, where $m_{q}$ is the charged parton mass. Hence, the photon spectrum for the charged parton becomes prominent at LHC energies. The numerical results of our calculations for large-$p_{T}$ $\eta_{Q}$ mesons produced by the hard photoproduction gluon-gluon processes in relativistic heavy ion collisions are plotted in Fig.\,\ref{fg1} and Fig.\,\ref{fg2}.

In Fig.\,\ref{fg1} (Fig.\,\ref{fg2}), we plot the contributions from the hard photoproduction gluon-gluon processes to the $\eta_{c}$ ($\eta_{b}$) meson at mid-rapidity in p-p and Pb-Pb collisions at LHC energies. Compared with the production of the initial gluon-gluon interaction (the dashed line), the contribution of $\eta_{c,b}$ meson produced by semielastic hard photoproduction g-g processes (the dotted line) is not prominent in p-p collisons with $\sqrt{s_{NN}}=7.0\,\mathrm{TeV}$ and $\sqrt{s_{NN}}=14.0\,\mathrm{TeV}$, but the contribution of inelastic photoproduction g-g processes (the dashed-dotted line) becomes evident in p-p collisons [see Figs.\,\ref{fg1} (a), \ref{fg1} (b), \ref{fg2} (a), and \ref{fg2} (b)]. Indeed, for Pb-Pb collisions with $\sqrt{s_{NN}}=2.76\,\mathrm{TeV}$ and $\sqrt{s_{NN}}=5.5\,\mathrm{TeV}$, the contribution of semielastic photoproduction g-g processes (the dotted line) and inelastic photoproduction g-g processes (the dashed-dotted line) cannot be negligible at LHC energies [see Figs.\,\ref{fg1} (c), \ref{fg1} (d), \ref{fg2} (c), and \ref{fg2} (d)].

\section{\label{sec:Conclusions}Conclusions}

In summary, we have investigated the production of heavy quarkonium $\eta_{c,b}$ meson from the gluon-gluon interactions in p-p collisions and Pb-Pb collisions at LHC energies. The color singlet and color octet mechanisms have been used for heavy quarkonium production processes. At the early stages of relativistic high energy nucleus-nucleus collisions, the ultrarelativistic nucleus (charged parton) can emit hadron-like photons that can fluctuate into a gluon, then the gluon interacts with a gluon of the other incident nucleus by gluon-gluon interaction. Our results indicate that the contribution of $\eta_{c,b}$ meson produced by the hard photoproduction processes cannot be negligible in p-p and Pb-Pb collisions at LHC energies.

\section*{\label{sec:Ack}Acknowledgments}
This work has been supported by the National Basic Research Program of China (973 Program, 2014CB845405), the China Postdoctoral Science Foundation funded project (2017M610663), and the Applied Basic Research Plan of Yunnan Province (Youth Project, 2017FD147).


%

\end{document}